# Persistent Neutron Irradiation Effects and Partial Recovery in Thin-Film CdS/CdTe Solar Cells

Ashraful Mamun[1], Andrew Allison[2], Meng-Jen (Vince) Wang[2], Etee Kawna Roy[1], Edward Cazalas[2], Glenn E. Sjoden[2], Matthew O. Reese[3], Joel N. Duenow[3], and Heayoung P. Yoon[1]*

[1] Department of Electrical and Computer Engineering, University of Utah, Salt Lake City, Utah 84112, USA

[2] Utah Nuclear Engineering Program, Department of Civil and Environmental Engineering, University of Utah, Salt Lake City, Utah 84112, USA

[3] National Laboratory of the Rockies, Golden, Colorado 80401, USA

Thin-film CdTe solar cells are promising candidates for lightweight space power, but device-level studies of their response to neutron irradiation remain scarce. Here, we report the device performance of CdS/CdTe solar cells irradiated for 10 to 150 min in the thermal irradiator of a TRIGA reactor (90 kW), corresponding to nominal total neutron fluences of $3.3 \times 10^{14}$ to $5.0 \times 10^{15}$ $n \cdot cm^{-2}$. The cells were characterized after 18 months of post-irradiation storage, once the residual activity has decayed to background levels. Co-irradiated FTO-coated glass reference substrates exhibit a wavelength-dependent transmission loss, with the transmittance at 500 nm decreasing from 80 % to 50 % after 60 min of irradiation. The irradiated cells retain rectifying behavior, but the reverse-bias dark current, photocurrent, and efficiency generally decrease with exposure time. Post-irradiation annealing at 70 °C under one-sun illumination initially improves $J_{sc}$ and $V_{oc}$, with the highest efficiency after 4 h. Prolonged annealing, however, causes substantial *FF* degradation despite continued photocurrent recovery. These trends suggest that irradiation-induced changes in the glass substrate, CdTe absorber, and contacts respond differently to annealing. Understanding underlying defect chemistry and correlating it with device behavior are necessary steps toward radiation-tolerant CdTe photovoltaics for space applications.

*Corresponding Author: heayoung.yoon@utah.edu

## 1. Introduction

Photovoltaic (PV) technology is an established means of power generation for space applications. III-V multijunction solar cells are a leading architecture for space solar power owing to their high conversion efficiency and radiation tolerance. However, their high production costs and the need for lower-mass power systems motivate the search for alternative PV technologies [1, 2]. Recent research efforts have focused on developing advanced silicon (Si) and organic solar cells, thin-film solar cells, and metal-halide perovskite single-junction and tandem architectures [3, 4]. Among them, cadmium telluride (CdTe) solar cells are promising candidates, combining thin absorber layers (a few µm thick) with manufacturing processes already established for cost-effective and high-performance terrestrial PV [5, 6] .

Previous studies have demonstrated the tolerance of CdTe-based solar cells to charged-particle irradiation. Om accelerated irradiation tests, Cu- and As-doped CdSeTe cells retained about 95 % and 80 %, respectively, of the efficiency of unirradiated controls after exposure to 650 keV protons at a fluence of $10^{12}$ $cm^{-2}$ [7]. Related modeling predicted that CdSeTe cells would retain a grater fraction of their initial efficiency than the III-V multijunction cells evaluated in a simulated medium Earth orbit (MEO) environment [8]. Over six years in orbit, CdTe cells on ultra-thin cesium-doped cover glass aboard the AlSat-1N CubeSat showed no decline in short-circuit current density ($J_{sc}$). The authors attributed the decline in FF declined to Au diffusion from the back contact [9, 10].

While such results support considering CdTe PV technology for a broad range of space environments, mission-specific assessments must also consider secondary radiation and radiation sources in the surrounding environment. Unlike charged particles (i.e., protons and electrons), neutrons and gamma-ray photons are weakly attenuated by cover glass and thin shielding. Neutrons can cause displacement damage and ionization in solar cells. Secondary neutrons are often generated when energetic particles interact with planetary surfaces, atmospheres, and surrounding materials [11]. For PV arrays operating near fission power systems, as proposed for lunar surface installations, reactor leakage radiation may introduce additional, localized neutron and gamma-ray exposure that depends on shielding, separation distance, and installation geometry [12].

CdTe presents a material-specific susceptibility to thermal-neutron capture, because the naturally occurring isotope $^{113}Cd$ (≈ 12 % abundance) has an absorption cross-section of approximately 20,600 barns for 2,200 $m \cdot s^{-1}$ neutrons [13]. Even when averaged over natural isotopic abundances, Cd has an absorption cross-section of approximately 2,520 barns, far exceeding those of the constituent elements of other PV absorbers: Si (0.171 barns), Ga (2.75 barns), As (4.5 barns), Pb (0.171 barns), and I (6.15 barns), all at the same neutron speed [13]. Thermal-neutron capture by $^{113}Cd$ produces an excited $^{114}Cd$ nucleus, whose subsequent de-excitation emits a cascade of prompt gamma rays and imparts recoil to the residual nucleus through momentum conservation [14]. These processes may contribute to materials modification, including recoil-induced lattice-atom displacement. Early studies reported neutron-induced changes in the electrical and luminescence properties of CdTe crystals, and neutron damage has been examined in CdTe and CdZnTe radiation detectors [15, 16]. However, such studies do not resolve how neutron-induced changes in the thin-film CdTe absorber layer, interfaces, junction, and contacts collectively affect PV performance.

This study examines persistent optical and electrical changes in CdS/CdTe solar cells following reactor neutron irradiation and subsequent storage. Dark and light current-voltage (*I-V*) measurements are combined with optical characterization of co-irradiated fluorine-doped tin oxide (FTO)-coated glass reference substrates, and recovery is evaluated during annealing at 70 °C under one-sun illumination. Optical and electrical measurements are used to assess irradiation-induced optical attenuation in the substrate and changes in device performance, and the balance between recovery and degradation during annealing.

## 2. Experimental Section

### 2.1 Sample preparation

CdS/CdTe solar cells were supplied by the National Renewable Energy Laboratory (NREL; now the National Laboratory of the Rockies). The fabrication procedure is described elsewhere [17, 18]. Briefly, an *n*-type CdS buffer layer (≈ 100 nm) was sputtered onto an FTO-coated glass substrate (≈ 3 mm; TEC-12D), and a *p*-type CdTe layer (≈ 3.5 μm) was deposited by close-spaced sublimation, followed by sequential post-deposition treatments with $CdCl_2$ and $CuCl_2$. A thin Au layer (≈ 100 nm) was deposited on the CdTe by thermal evaporation as the back contact. Twelve cells from the same fabrication batch were divided into four irradiation groups of three cells each. Eight additional solar cells were retained as unirradiated controls. ***Figure 1a*** shows a representative cross-sectional scanning electron microscope (SEM) image, with characteristic CdTe grain dimensions on the order of 1 μm.

### 2.2 Neutron irradiation and storage

The solar cells were irradiated for 10, 30, 60, and 150 min in the dry thermal irradiator (TI) of the TRIGA reactor at the University of Utah, which was operated at 90 kW (***Figure 1b***). The cells were mounted on cardstock strips inside polyethylene vials (inset of ***Figure 1b***) and suspended in the chamber using a plastic thread (≈ 6 m long). The same positioning arrangement was used for each group. The neutron flux distribution of the reactor, including the TI, has been modeled elsewhere [19]. The sample temperature during irradiation was not measured.

The reported neutron-energy range in the TI chamber was 0.02 eV to 20 MeV, with estimated fast and thermal neutron fluxes of $1.5 \times 10^{11}$ and $4.0 \times 10^{11}$ $n \cdot cm^{-2} \cdot s^{-1}$, respectively. The nominal total neutron fluences, calculated from the sum of the fast and thermal fluxes, were $3.3 \times 10^{14}$, $9.9 \times 10^{14}$, $2.0 \times 10^{15}$, and $5.0 \times 10^{15}$ $n \cdot cm^{-2}$ for irradiation times of 10, 30, 60, and 150 min, respectively (***Table S1***, Supporting Information). The samples in the TI were also exposed to the gamma field of the reactor. Neutron- and gamma-induced effects are not separated in this work.

After irradiation, the solar cells and optical references (***Section 2.3***) were stored in lead-shielded containers in the dark. Residual count rates were monitored using a NaI scintillation detector coupled to a Ludlum 2241-2 survey meter. In accordance with the facility safety protocol, the cells were characterized after 18 months of post-irradiation storage, once the residual activity had decayed to background levels. Except during characterization and the annealing treatment, the samples remained in the dark. The results therefore reflect persistent rather than transient responses. Any recovery that may have occurred during storage was not monitored.

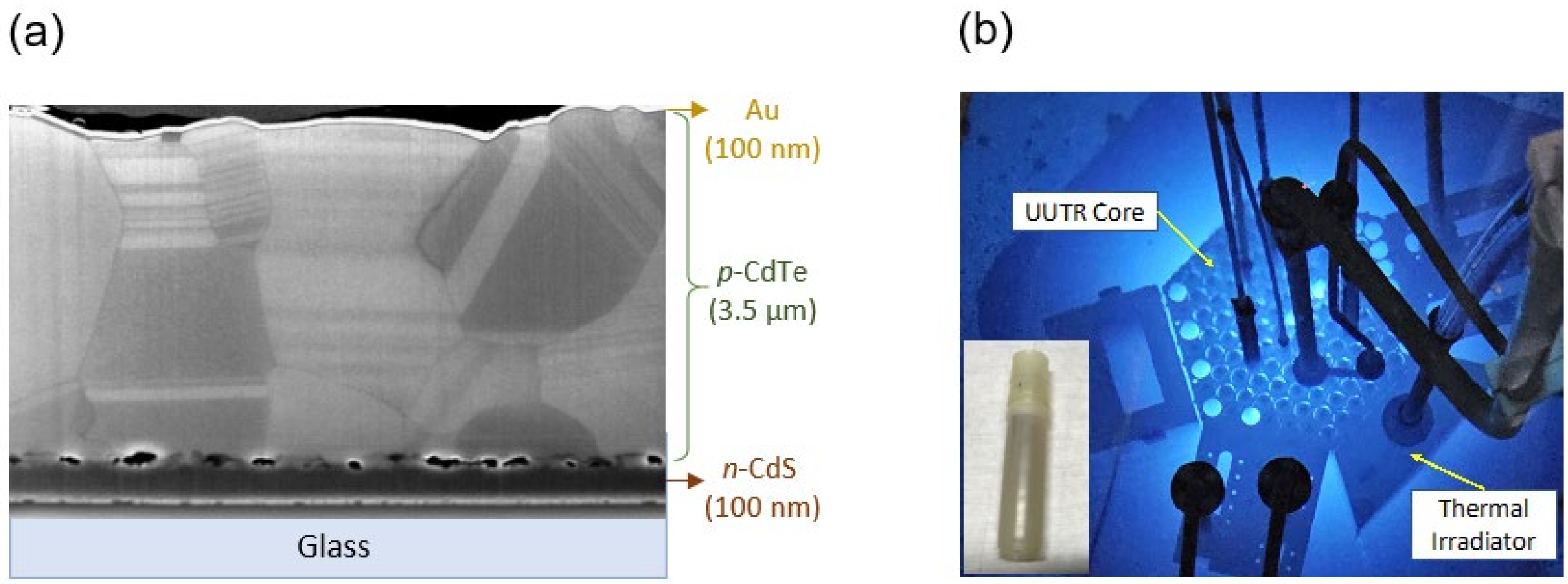

***Figure 1.*** (a) Schematic of a CdS/CdTe solar cell before neutron irradiation, showing a cross-sectional SEM image on top of a glass substrate (not to scale). (b) Location of the thermal irradiator (TI) in the TRIGA reactor at the University of Utah. The inset shows the configuration of the sample mounting vial (≈ 25 mm diameter). The reactor was operated at 90 kW.

### 2.3 Optical characterization

Separate FTO-coated glass references (≈ 2 mm; TEC-7) were also mounted and co-irradiated with the CdTe solar cells for 10, 30, and 60 min. Transmittance (T) and reflectance (R) were measured over 300-1100 nm in 2 nm increments using UV-Vis-NIR spectrophotometer (PerkinElmer Lambda 950). The samples were illuminated from the glass side through an aperture (4 mm diameter). Transmittance was measured at normal incidence using an integrating sphere, while reflectance was measured at 8° using a universal reflectance accessory.

### 2.4 Electrical characterization

Dark and light *I-V* curves were obtained using a probe station and a semiconductor parameter analyzer (Agilent 4156C) with a voltage sweep from -1 V to +1 V in 10 mV increments. Light measurements used a solar simulator (G2V) calibrated to nominal one-sun illumination (AM 1.5G) with a reference Si cell (Newport 91150V). The short-circuit current density ($J_{sc}$), open-circuit voltage ($V_{oc}$), fill factor (*FF*), and power conversion efficiency were extracted from the light *I-V* curves. Additional LED-based spectral response measurements are described in ***Section S3***, Supporting Information.

### 2.5 Annealing

The recovery experiment was performed on a cell irradiated for 150 min. The device was heated on a hot plate at 70 °C under one-sun illumination (light soaking), and *I-V* curves were obtained after cumulative treatment times of 2, 4, 6, 8, and 10 h. An additional measurement was performed after 7 days of dark storage following the annealing sequence.

## 3. Results and Discussion

### 3.1 Irradiation-induced optical changes

Progressive darkening is evident in the side-view photographs of the solar cells after neutron irradiation (***Figure 2a***). The substrate becomes pale brown to amber after 10-30 min of irradiation and appears nearly black after 60-150 min. The optical references (FTO-coated glass; TEC-7) also develop a brown coloration (***Figure 2b***), demonstrating that the optical changes occur in the FTO-coated glass, independently of the CdS/CdTe layers. The different viewing geometries (the CdTe cells are photographed edge-on), the thicker device substrate (≈ 2 mm vs. ≈ 3 mm), and the presence of the solar cell stack likely contribute to the darker appearance of the CdTe solar cells relative to the optical references.

The optical changes are strongly wavelength dependent. The near-ultraviolet reflectance peak around 325 nm decreases from approximately 11 % to 6 % after 60 min, whereas the spectra are closely comparable above 500 nm (***Figure 2c***). Transmittance decreases substantially across much of the ultraviolet-visible range (***Figure 2d***). At 500 nm, the approximate values are 80 %, 70 %, 60 %, and 50 % for the unirradiated, 10 min, 30 min, and 60 min samples, respectively. The differences diminish toward longer wavelengths, and the curves converge near 800 nm. The transmission loss without increased reflectance indicates greater absorption in the FTO-coated glass substrate.

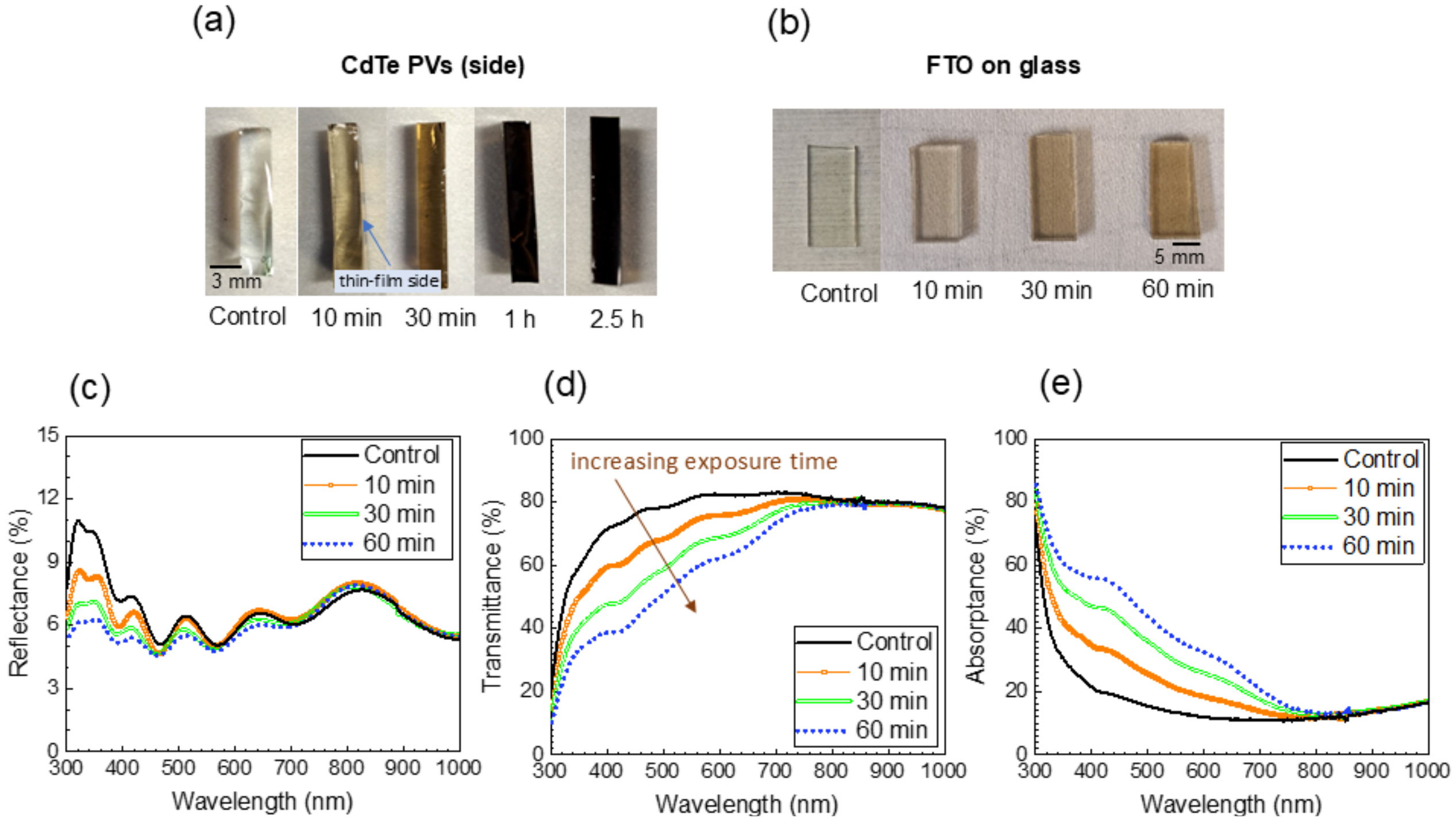

***Figure 2.*** Irradiation-induced optical changes. (a) Side-view photographs of CdS/CdTe solar cells on ≈ 3 mm substrates (TEC-12D) in the unirradiated state and after 10, 30, 60, and 150 min of irradiation. (b) Photographs of the ≈ 2 mm optical references (TEC-7) in the unirradiated state and after 10, 30, and 60 min of irradiation. Viewed edge-on, these references show color changes similar to those in (a). (c) Reflectance (R), (d) transmittance (T), and (e) absorptance (A = 1 - T - R) of the optical references. "Control" denotes the unirradiated sample.

***Figure 2e*** shows the estimated absorptance (A = 1 - T - R). The additional photon loss due to absorption in the glass, before light enters the CdS/CdTe layers is approximately 10 %, 21 %, and 28 % of the incident photons at 500 nm after 10 min, 30 min, and 60 min, respectively, and the corresponding values at 400 nm are approximately 13 %, 26 %, and 34 %. The induced absorptance increases toward shorter wavelengths and grows sublinearly with irradiation time. A sixfold increase in irradiation time raises the absorptance change by a factor of only about three. Sublinear growth with dose is commonly observed for radiation-induced color centers in silica-based glasses [20]. Because this radiation-induced absorption lies within the spectral response range of CdTe (absorption edge near 855 nm; $E_g \approx 1.45$ eV), it contributes to the photocurrent loss. Weighting the non-absorbed fraction, 1 - A(λ), where λ is the wavelength, by the AM 1.5G photon flux, we estimate that the glass darkening alone reduces $J_{sc}$ by approximately 7 %, 14 % and 20 % relative to the control after 10 min, 30 min, and 60 min of irradiation, respectively.

These observations are consistent with our previous report on glass-side photoluminescence (PL), which showed enhanced broad emission near 640 nm associated with radiation-induced color centers [21]. Emission in this region is consistent with non-bridging oxygen hole centers (NBOHCs) in silica-based materials. In soda-lime silicate glass, NBOHCs are the main source of the brown coloration produced by ionizing radiation, with induced absorption bands near 430 nm and 630 nm [22]. Such coloration fades only partly at room temperature and can persist for months [22], as observed in our samples after 18 months of storage. The TI chamber delivers a mixed neutron and gamma field. Gamma-induced discoloration and increased optical absorption have been reported in soda-lime silicate glass [23], suggesting that the reactor gamma field is a plausible contributor to the darkening observed in our samples, in addition to prompt gamma rays generated by neutron interactions within the samples. The present data cannot establish the relative contributions of these components. Because the induced absorption scales with glass thickness, the solar cell substrates (≈ 3 mm TEC-12D) should attenuate more strongly than the optical references (≈ 2 mm TEC-7), suggesting that the reference spectra provide a lower bound on the optical loss in the solar cells.

### 3.2 Photovoltaic response to irradiation

***Figures 3a*** and ***3b*** compare representative dark and light *I-V* curves. The data are compared with the baseline performance of the unirradiated control cells. The eight baseline devices had a mean efficiency of 11 ± 2.2 %, $J_{sc}$ of 24.4 ± 3.3 mA/cm$^2$, $V_{oc}$ of 0.83 ± 0.02 V, and $FF$ of 0.54 ± 0.05. The cells retain rectifying behavior after all irradiations. In reverse bias, the magnitude of the dark current at -1 V is approximately $1.8 \times 10^{-8}$ A before irradiation and approximately $1.3 \times 10^{-9}$ A after 150 min, although the decrease is not monotonic at the shortest irradiation times.

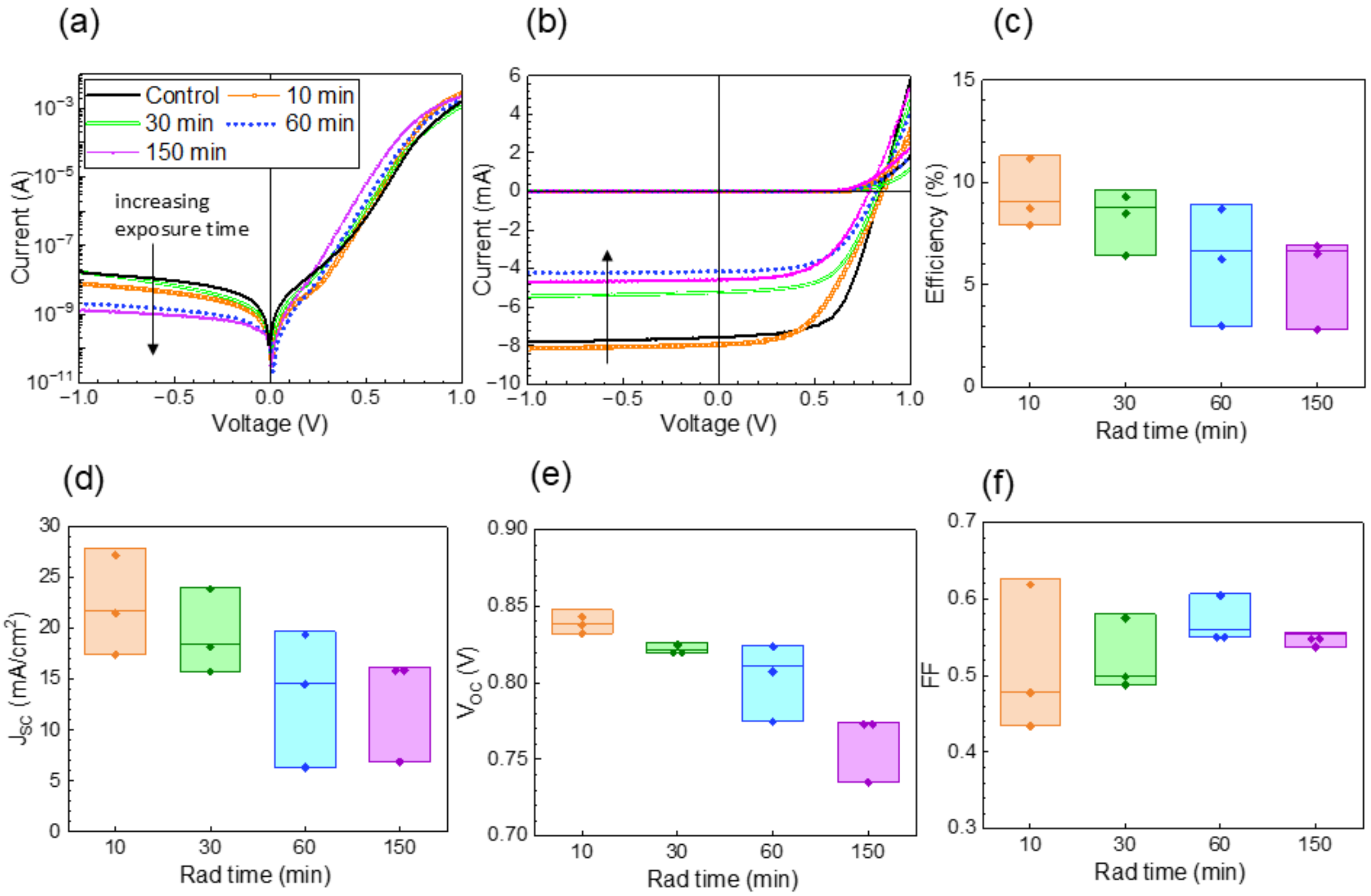

***Figure 3.*** Photovoltaic response after irradiation and post-irradiation storage. Representative (a) dark and (b) light current-voltage (*I-V*) curves. Distributions of (c) efficiency, (d) $J_{sc}$, (e) $V_{oc}$, and (f) *FF* for the three solar cells in each irradiation group. "Control" denotes the unirradiated cells. Nominal total neutron fluences for 10, 30, 60, and 150 min are $3.3 \times 10^{14}$, $9.9 \times 10^{14}$, $2.0 \times 10^{15}$, and $5.0 \times 10^{15}$ n·cm$^{-2}$, respectively, calculated from the nominal fast and thermal fluxes of the thermal irradiator.

Under illumination, the representative 10 min cell retains a $J_{sc}$ comparable to that of the unirradiated cell, whereas cells irradiated for longer exhibit lower current magnitudes (***Figure 3b***). The light *I-V* curves remain nearly flat between 0 V and -1 V for all irradiation times, indicating that the lost photocurrent is not recovered by reverse bias. A deficit that does not respond to the collecting electric-field suggests reduced photogeneration rather than incomplete carrier collection, consistent with the substrate attenuation described in ***Section 3.1***. In contrast, the lower dark current after longer irradiation implies an altered electrical response that reflects combined contribution is a lower free-carrier (hole) concentration in the *p*-type CdTe caused by compensating defects and complexes, as reported by Barnes and Kikuchi for thermal-neutron irradiated CdTe crystals [15, 24].

***Figures 3c-f*** and ***Table 1*** summarize the PV parameters for the three cells in each irradiation group. The distributions exhibit noticeable cell-to-cell variation. The 10 min $J_{sc}$ values overlap the range of baseline performance (24.4 ± 3.3 mA/cm$^2$). The longer-irradiation groups generally show lower $J_{sc}$. After 150 min of irradiation, the individual values span 6.9~16.2 mA cm$^{-2}$ (***Figure 3d***). The group mean $J_{sc}$ values correspond to approximately 92 %, 79 %, 55 %, and 53 % of the baseline mean after 10, 30, 60, and 150 min of irradiation, respectively. Most of the photocurrent loss develops within the first 60 min, broadly consistent with the sublinear growth of the induced absorptance (***Section 3.1***). The LED-based spectral response of a cell irradiated for 150 min is consistent with a largely optical origin of the photocurrent loss (***Figure S2***, Supporting Information). Relative to the pre-irradiation response, the loss is largest at short wavelengths. The irradiated cell retains only roughly 10-30 % of the initial response between 400 nm and 570 nm but about 45-65 % between 690 nm and 820 nm.

The 150 min group also has a lower $V_{oc}$, approximately 0.73-0.77 V, than the shorter-irradiation groups (***Figure 3e***). The 10 min group lies slightly above the unirradiated median of approximately 0.83 V but within the baseline spread, while the 30 min group remains close to it. In contrast to $J_{sc}$ and $V_{oc}$, *FF* does not decrease monotonically (***Figure 3f***). The 10 min group has the widest spread, and the 30-150 min values are broadly comparable and remain within the baseline range (0.54 ± 0.05). The power conversion efficiency is generally lower for the longer irradiations (***Figure 3c***), reflecting the reduced photocurrent and the lower voltage rather than a systematic *FF* decline. We note that the trends observed here should be regarded as indicative rather than statistically resolved, given the three cells per group. Nevertheless, the optical and electrical measurements identify two contributions. Front-substrate optical attenuation reduces the light available for photogeneration and accounts for a large part of the photocurrent loss, while the changes in the dark characteristics and the lower $V_{oc}$ indicate electrical modification.

**Table 1.** Photovoltaic parameters after irradiation and post-irradiation storage. Values are means, which ranges in parentheses, for three cells per irradiation group and for the eight unirradiated cells (baseline).

| Irradiation Time (min) | Total Neutron Fluences ($n \cdot cm^{-2}$) | $J_{SC}$ ($mA/cm^2$) | $V_{OC}$ (V) | FF (%) | Efficiency (%) |
|---|---|---|---|---|---|
| Baseline | | 24.4 (17.0~ 27.4) | 0.83 (0.78 ~ 0.84) | 54.1 (45.1 ~ 60.4) | 11.0 (6.7 ~ 13.5) |
| 10 min | $3.3 \times 10^{14}$ | 22.4 (17.4 ~ 27.9) | 0.84 (0.83 ~ 0.85) | 51.2 (43.4 ~ 62.7) | 9.4 (7.9 ~ 11.3) |
| 30 min | $9.9 \times 10^{14}$ | 19.4 (15.7 ~ 24.0) | 0.82 (0.82 ~ 0.83) | 52.2 (48.8 ~ 58.0) | 8.3 (6.4 ~ 9.7) |
| 60 min | $2.0 \times 10^{15}$ | 13.6 (6.3 ~ 19.7) | 0.80 (0.77 ~ 0.82) | 57.2 (55.0 ~ 60.7) | 6.2 (3.0 ~ 8.9) |
| 150 min | $5.0 \times 10^{15}$ | 13.1 (6.9 ~ 16.2) | 0.76 (0.73 ~ 0.77) | 54.9 (53.7 ~ 55.7) | 5.5 (2.8 ~ 6.9) |

### 3.3 Neutron capture, recoil energy, and possible defect mechanisms

The large thermal-neutron absorption cross-section of $^{113}$Cd (***Table S2***; Supporting Information) provides a plausible route to defect formation in the Cd-containing layers, including the CdS window. Capture produces an excited $^{114}$Cd* nucleus, which returns to the ground state through a prompt gamma-ray cascade ($\sum_i \gamma_i$) with a total energy release of approximately 9.04 MeV [25].

$$^{113}_{48}\mathrm{Cd} + {}^{1}_{0}\mathrm{n} \rightarrow ({}^{114}_{48}\mathrm{Cd})^{*} \rightarrow {}^{114}_{48}\mathrm{Cd} + \textstyle\sum_i \gamma_i \qquad (1)$$

Each photon ($\gamma_i$) emitted during the cascade imparts recoil momentum to the $^{114}$Cd nucleus. For a single photon of energy $E_\gamma$ emitted by a nucleus of mass M initially at rest, momentum conservation gives the recoil energy in the nonrelativistic, free-nucleus approximation:

$$E_R \approx \frac{E_\gamma^2}{2Mc^2} \qquad (2)$$

Here, $c$ is the speed of light. The nucleus recoils in the opposite direction to the emitted gamma ray, and the recoil energy increases with the square of the gamma-ray energy. When several gamma rays are emitted one after the other, the net recoil also depends on their relative directions and on how far the nucleus moves between emissions. The $^{114}$Cd* nucleus formed by neutron capture ($M \approx 114$ u; $Mc^2 \approx 1.06 \times 10^5$ MeV) de-excites to $^{114}$Cd by emitting approximately four gamma rays on average [26]. About one of the four belongs to the three strongest lines, at 558, 651, and 806 keV, while the other three are distributed over many weaker lines, mostly at higher energies [27, 28] (***Section S2***; Supporting Information).

Whether this recoil damages the CdTe absorber layer depends on the threshold displacement energy ($E_d$), the minimum kinetic energy that an atom must receive to be knocked out of its lattice site. The displaced atom and the empty site form a Frenkel pair (a Cd vacancy [$V_{Cd}$] and a Cd interstitial [$Cd_i$]). We adopt $E_d \approx 5.6$ eV for Cd and $E_d \approx 7.9$ eV for Te in CdTe from a previous study [29]. With this Cd threshold, Equation (2) shows that a single gamma ray must carry at least 1.09 MeV to displace the Cd atom. The three strongest lines (558, 651, and 806 keV) impart only 1.5, 2.0, and 3.1 eV, respectively, and cannot displace the atom on their own. However, most of the remaining gamma rays exceed this energy (e.g., 3 MeV photon imparts about 42 eV), and the recoil from one of them is sufficient to displace the Cd atom that captured the neutron and, in many cases, a few neighboring atoms [24]. Donor-like Cd interstitials or complexes are candidate compensating defects in *p*-type CdTe, whereas Cd vacancies are acceptor-like defects that can also facilitate non-radiative recombination [30]. The formation of a vacancy-interstitial pair does not by itself determine the net free-carrier change, because the two members of the Frenkel pair have in opposite electronic effects and may recombine, migrate, or form complexes with Cu and Cl.

The TI chamber in the TRIGA reactor contains a spectrum of both fast and thermal neutrons, with fluxes estimated at $1.5 \times 10^{11}$ and $4.0 \times 10^{11}$ n·cm$^{-2}$·s$^{-1}$, respectively. Fast neutrons with energies in the keV-MeV range can transfer up to approximately 3.5 % of their energy to a Cd atom in a single elastic collision and can produce displacement cascades [24]. The gamma-ray field also generates energetic secondary electrons, which can cause atomic displacement at sufficiently high energies. The relative contributions of thermal-neutron capture recoil, fast-neutron collisions, and gamma-ray-induced secondary electrons to atomic displacement cannot be determined from the present study. Nevertheless, Barns and Kikuchi reported that thermal-neutron capture recoil damage exceeded fast-neutron damage in CdTe under the conditions of their reactor experiments [15]. We therefore speculate that thermal-neutron capture recoil may be the dominant contributor to the electrical changes observed in the cells.

### 3.4 Partial recovery during low-temperature annealing

***Figure 4*** shows the annealing response of a cell irradiated for 150 min. Side-view photographs show partial bleaching after 10 h at 70 °C under one-sun illumination, although a yellow-brown coloration remains (***Figure 4a***). This is qualitative evidence that some of the visible darkening is reversible, as expected for radiation-induced color centers in glass, which bleach thermally and optically [23].

The light *I-V* curves show increasing $J_{sc}$ during the annealing sequence ***(Figure 4b***). $J_{sc}$ rises from approximately 16 mA/cm$^2$ before annealing to approximately 23 mA/cm$^2$ after 10 h (***Figure 4d***). $V_{oc}$ initially increases from approximately 0.77 V to a maximum near 0.83 V at 6 h, followed by a small decline (***Figure 4e***). This gain of about 60 mV brigs $V_{oc}$ back to the mean of the unirradiated cells (0.83 ± 0.02 V). ***FF*** remains approximately 0.54 up to 4 h (***Figure 4f***), so the early increases in current and voltage improve the efficiency, which reaches its highest measured value at 4 h (***Figure 4c***). At that point the efficiency has recovered from about 60 % (post-irradiation) to about 82 % (4 h annealing) of the average value of the baseline control. Beyond 4 h, *FF* decreases to about 0.38 at 6 h and 0.27 at 8-10 h. This loss outweighs the continued photocurrent recovery, and the efficiency after 10 h falls below the post-irradiation value (before annealing).

The ***FF*** decrease coincides with increasing distortion of the light *I-V* curves (***Figure 4b***). Up to 4 h, the curves keep a flat reverse-bias plateau and a steep rise beyond $V_{oc}$. From 6 h onward, two distortions appear. First, the forward current beyond $V_{oc}$ is strongly suppressed. At +1 V, it falls from roughly 4~5 mA after 2~4 h to less than 2 mA after 6~8 h and less than 1 mA after 10 h. Such roll-over is the characteristic signature of a current-limiting barrier at the back contact. Second, the photocurrent no longer saturates in reverse bias. After 10 h, it becomes clearly voltage-dependent carrier collection. Such behavior is compatible with contact degradation, for example through Cu redistribution and/or Au migration, which can modify the CdTe back contact under combined heat and illumination [31]. The treatment used in this study, however, is mild compared with the conditions normally used to stress unirradiated CdTe cells, which may indicate that irradiation made the contact region more susceptible to such changes. After 7 days of subsequent dark storage. $V_{oc}$ remains close to its 10 h value, but $J_{sc}$ and *FF* decrease slightly further, so the degraded state does not relax in the dark on this timescale.

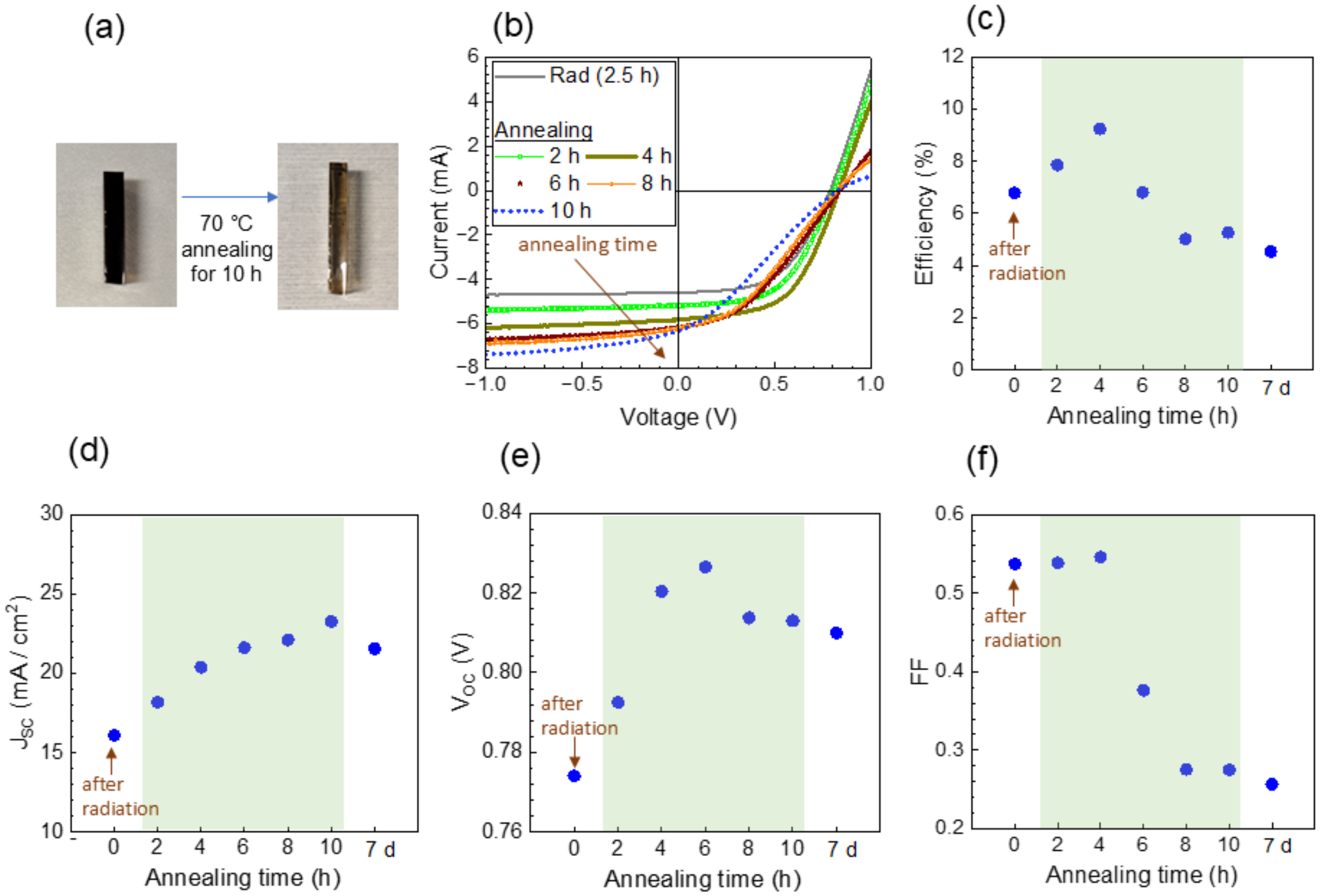

***Figure 4.*** Annealing response of a cell irradiated for 150 min. (a) Side-view photographs showing partial bleaching of the glass after annealing at 70 °C for 10 h. (b) Light *I-V* curves of the cell before and after irradiation and at different times during annealing at 70 °C under one-sun illumination. Evolution of the photovoltaic parameters of the same cell with annealing and after 7 days in the dark: (c) power conversion efficiency, (d) short-circuit current density ($J_{SC}$), (e) open-circuit voltage ($V_{OC}$), and (f) fill factor (*FF*).

Annealing-related recovery is consistent with earlier observations in irradiated CdTe, although the protocols and defect populations differ. Barnes and Kikuchi reported recovery of neutron-induced electrical changes at about 160~200 °C using short isochronal treatments [15], while Lamb *et al.* observed recovery after proton irradiation using 100 °C annealing for 168 h in nitrogen [32]. Thermal recovery of radiation damage has also been reported for CdTe and CdZnTe detectors [33], and post-irradiation thermal or heat-light treatments recover part of the damage in III-V solar cells [34]. The 70 °C used here is well below the temperatures at which the neutron-induced electrical changes in bulk CdTe were reported to anneal, so the partial $V_{oc}$ recovery probably involves only the least stable defects. Vacancy-interstitial recombination is one possible pathway, but optical bleaching, changes in recombination activity, and contact evolution may occur on different timescales.

### 3.5 Implications and limitations

The reactor irradiations used in this study are accelerated tests rather than a simulated of a specific space environment. Secondary neutron fluxes at the lunar and Martian surfaces are many orders of magnitude lower than the flux in the TI chamber [35, 36]. For example, Monte Carlo calculations give a cosmic-ray-induced neutron leakage flux of about 2~4 $n \cdot cm^{-2} \cdot s^{-1}$ at the lunar surface (roughly $10^8$ $n \cdot cm^{-2}$ per year). At this rate, accumulating $10^{14}$~$10^{15}$ $n \cdot cm^{-2}$ would take $10^6$~$10^7$ years, so the present irradiations are best interpreted as high-fluence test conditions. In contrast, their relevance to solar arrays operated close to fission power sources depends on reactor power, shielding, separation distance, and exposure duration. Design studies for lunar fission surface power systems predict neutron fluences in this range for hardware mounted on the same lander as the reactor. For example, the control-electronics box of a shielded 500 kWth lunar reactor is estimated to receive $10^{14}$~$10^{15}$ $n \cdot cm^{-2}$ per year, and the radiator $10^{15}$~$10^{17}$ $n \cdot cm^{-2}$ per year [12]. The fluences for reactor-adjacent solar panels in this work correspond to 0.3~50-years at the electronic-box location, depending on the rate. We note that this comparison is on the fluence basis, as the TI used in this work is predominantly thermal neutrons.

Under the conditions tested, our results suggest two perspectives. First, much of the photocurrent loss is associated with darkening of the soda-lime glass substrate, a partly reversible optical effect that originated outside the CdTe absorber. Cerium-doped cover glass, designed to resist radiation darkening and already demonstrated in CdTe solar cells [35], is a promising candidate substrate for future neutron tests [32]. Second, the lower $V_{oc}$, changes in the dark I-V characteristics, and loss of *FF* during prolonged annealing point to electrical changes in the CdTe absorber and contacts. The decline in FF during annealing suggests that contact stability may limit recovery. Further studies are needed to develop more stable back contacts of CdTe-based solar cells for space applications.

The present work captures the combined effect of neutron and gamma irradiation. Comparing gamma-only controls with and without thermal-neutron filtering would help distinguish the contributions of gamma rays, thermal neutrons, and fast neutrons. Pairing these comparisons with optical and electrical measurements sensitive to absorber and contact properties could identify the source of the performance loss. *In-situ* measurements during irradiation and time-resolved post-irradiation measurements would help distinguish immediate radiation-induced changes from subsequent recovery during storage.

## 4. Conclusion

Neutron irradiation in a TRIGA reactor, at nominal total fluences of of $3.3 \times 10^{14}$ to $5.0 \times 10^{15}$ $n \cdot cm^{-2}$, produces optical and electrical changes in CdS/CdTe solar cells that persist after 18 months of storage. The cells retain rectifying behavior but generally exhibit lower photocurrent and efficiency at longer irradiation times. Darkening of the front glass substrate reduces transmission and accounts for a large part of the photocurrent loss, while the modified dark *I-V* characteristics and the lower $V_{oc}$ after the longest irradiation (150 min) indicate additional electrical changes. The recoil energy of $^{114}$Cd following thermal-neutron capture can exceed the Cd displacement threshold of about 5.6 eV, which makes recoil-induced point defects a plausible contributor to these changes. Annealing at 70 °C under one-sun illumination initially improves performance, with the highest recovered efficiency reached after 4 h. Prolonged annealing instead causes substantial *FF* degradation despite continued photocurrent recovery. Mild annealing therefore provides only partial recovery. Further work is needed to quantify the loss contributions from the bulk, interfaces, contacts, and microstructure and to separate neutron- and gamma-induced effects.

**Acknowledgements**

We thank M. Beitollahi and the EHS staff members at the University of Utah for valuable help with radiation safety, and K. Watts for assistance with UV-Vis-NIR measurements. This work was partially supported by the U.S. National Science Foundation under CAREER Award No. 2048152 and by a 1U4U Seed Grant from the University of Utah. This work was authored in part by the National Laboratory of the Rockies for the U.S. Department of Energy (DOE), operated under Contracts No. DE-EE0008983 and DE-AC36-08GO28308. Funding provided by U.S. Department of Energy Office of Critical Minerals and Energy Innovation Integrated Energy Systems Office under project numbers 37989 and 52778. The views expressed in the article do not necessarily represent the views of the DOE or the U.S. Government. The U.S. Government retains and the publisher, by accepting the article for publication, acknowledges that the U.S. Government retains a nonexclusive, paid-up, irrevocable, worldwide license to publish or reproduce the published form of this work, or allow others to do so, for U.S. Government purposes.

**Conflict of Interests**

The authors declare no conflict of interest.

**Data Availability Statement**

The data that support the findings of this study are available from the corresponding author upon reasonable request.

**Declaration of Generative AI Technologies in the Manuscript Preparation Process**

The authors used ChatGPT (OpenAI) and Claude (Anthropic) to assist in finding relevant literature, improving sentence clarity, and correcting grammar. The authors reviewed and verified the contents and take full responsibility for the content of this article.